# Focusing Ultrasound with Acoustic Metamaterial Network

**Shu Zhang, Leilei Yin and Nicholas Fang**
*Department of Mechanical Science and Engineering, University of Illinois at Urbana- Champaign, USA*

We present the first experimental demonstration of focusing ultrasound waves through a flat acoustic metamaterial lens composed of a planar network of subwavelength Helmholtz resonators. We observed a tight focus of half-wavelength in width at 60.5 KHz by imaging a point source. This result is in excellent agreement with the numerical simulation by transmission line model in which we derived the effective mass density and compressibility. This metamaterial lens also displays variable focal length at different frequencies. Our experiment shows the promise of designing compact and light-weight ultrasound imaging elements.

High-resolution acoustic imaging techniques are the essential tools for nondestructive testing and medical screening. However, the spatial resolution of the conventional acoustic imaging methods is restricted by the incident wavelength of ultrasound. This is due to the quickly fading evanescent fields which carry the subwavelength features of objects. To overcome this diffraction limit, a remarkable perfect lens is proposed by John Pendry [1], which offers the promise to build a device allowing super-resolution imaging of an object. This perfect lens is based on focusing the propagating wave and recovering the evanescent field through a flat negative-index slab. Since then research on metamaterials has been stimulated by the opportunity to develop artificial media that refracts waves in negative direction. Several different metamaterials have been proposed and demonstrated to present negative index of refraction [2-6].

The successful demonstration of electromagnetic (EM) superlens [7-10] has inspired the search for the analogous acoustic negative-index lens. In fact, phononic crystals [11-15] were first investigated to develop negative-refractive devices for sound waves. Beam steering in phononic crystals can be achieved by Bragg scattering, leading to enhanced diffraction in negative direction. Ultrasound focusing from negative refraction by a three-dimensional phononic crystal was first demonstrated experimentally by Yang $et$ $al.$[13].A focal spot around five wavelengths in width was observed in the far field at 1.57 MHz. Recently, a finer resolution was achieved by focusing the ultrasound field emitted by a subwavelength line source using a two-dimensional (2D) phononic crystal slab[15].

However, for lens design based on phononic crystals, the dependence of band structure on the lattice periodicity usually requires the spatial modulation to be the same order of magnitude as the acoustic wavelength, which would makes such structure impracticably large. Locally resonant sonic materials [16] made a major step towards the acoustic metamaterial development. Since the lattice constant is much smaller than the relevant wavelength, effective medium properties can be attributed to this sonic material at low frequency. With appropriate resonances included into the building block, acoustic metamaterials with either negative effective mass density or bulk modulus or both have been demonstrated [17-20]. These anomalous phenomena resulted from strong coupling of the traveling elastic wave in the host medium with the localized resonance in the building block. However, to the best of our knowledge, there is no experimental demonstration of focusing ultrasound waves in these negative index acoustic metamaterials.

In this paper, we experimentally investigated the focusing of a point source from a designed ultrasonic metamaterial consisting of a planar network of subwavelength Helmholtz resonators. To facilitate the design, we adapted the 2D transmission line (TL) method which is widely used in the development of negative index EM metamaterials [8-10].In this approach, the acoustic system is converted to an analogous lumped circuit model in which the motion of the fluid is equivalent to the behavior of the current in the circuit. Similar to permittivity and permeability in the EM metamaterial [8], the effective density and compressibility of the network structure are found to be related to the capacitance and inductance in this lumped circuit. Earlier, in the one-dimensional version of this ultrasonic metamaterial, the elastic modulus is found to be negative at specific frequency range theoretically and experimentally [21].

Figure 1 shows the experimental setup to study the focusing phenomena of the acoustic metamaterial. To prepare the sample, we machined a 2D array of periodically connected subwavelength Helmholtz resonators in an aluminum plate and the resonators are filled with water. As shown in previous work [22-24], a main transmission channel with recurrent side branches, which are closed at the outer end, is analogous to a circuit of a series of inductors with shunt capacitors. On the other hand, when the side tubes inserted in the main channel is open on the outer end, the acoustic system can be described by a lumped network of a series of capacitors with shunt inductors. The left and right half parts in the sample are 2D periodic versions of those different types of topology respectively. One unit cell from each half part is enlarged and shown in the two insets respectively.

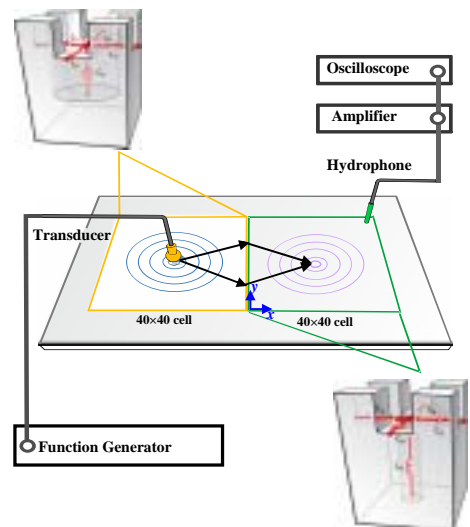

FIG. 1 (color). Schematic showing the experimental setup. The sample with PI/NI interface is composed of an array of different designed Helmholtz resonators machined from an aluminum plate. Unit cells of each half part and the corresponding inductor–capacitor circuit analogy are shown in the insets.

The left half part is composed of a 2D array (40×40) of larger cavities connected with main channels. The volume of the cavity is around ten times of that of one section of the channels. Consequently, when an incident acoustic wave is applied onto the fluid in the channels, the pressure gradient through the channels is much greater than that inside the cavity.

Hence, it is as if the fluid in the cavity were at rest relative to those in the channels [23]. So when the plug of fluid in the channels oscillates as a unit, there are adiabatic compressions and rarefactions of the fluid inside the larger cavity. Such an acoustic system is analogous to an inductor–capacitor circuit as shown in the inset with the channels acting as a series of inductors $(L_P)$ and the cavity providing the stiffness element as capacitors $(C_P)$. The periodicity (3.175mm) of the sample is one-eighth of the wavelength at around 60 KHz frequency range. Given this value, the lumped circuit model is a valid approximation for the distributed acoustic system with only 10% error [25]. Following the approach of EM circuit analysis [8-10], the effective density and compressibility of this network can be expressed in the form as $\rho_{eff,P} = \frac{L_P S_P}{d_P}$, $\beta_{eff,P} = \frac{C_P}{S_P d_P}$ where $d_P$ is the periodicity, $S_P$ is the cross section area of the channels. Both effective density and compressibility are positive. Effective relative acoustic refractive index $n_P$ can be determined by $n_P = \frac{c_w \sqrt{L_P C_P}}{d_P}$, where $c_w$ is speed of sound in water. We call this half part as effective positive index (PI) medium.

The right half part of the sample is the dual configuration of the left half part, in which there is an array (40×40) of orifices connected with channels. The volume of one section of the main channel is designed as around ten times of that of the orifice. Since the fluid in the orifice is not confined, it experiences negligible compression while the fluid in the channels experience less movement in average compared with that in the orifice [23]. Consequently, when the fluid in the orifice oscillates as a unit, there are adiabatic compressions and rarefactions of the fluid inside the main channels. Such an acoustic system is described as a lumped network with a series of capacitors $(C_N)$ for the main channel part and a shunt inductor $(L_N)$ due to the orifice. The periodicity is the same as that in the left part, so the effective mass density and compressibility can be similarly estimated as $\rho_{eff,N} = -\frac{S_N}{\omega^2 C_N d_N}$, $\beta_{eff,N} = -\frac{1}{\omega^2 L_N d_N S_N}$, where $d_N$ is periodicity and $S_N$ is the cross section area of connecting channels. Both parameters are negative. The refractive index $n_N = \frac{c_w}{v_\phi} = -\frac{c_w}{\omega^2 d_N \sqrt{L_N C_N}}$ is negative. So this network structure acts as a medium exhibiting negative index (NI) of refraction. The two half parts are designed with effective indices of equal and opposite value and matched impedance $\sqrt{\rho_{eff}/\beta_{eff}}$ at the design frequency 60.5 KHz.

For experimental confirmation of ultrasound focusing in this acoustic metamaterial, we measured the pressure field through this PI/NI interface. The ultrasound waves were launched from a horn shaped transducer with a tip of 3mm diameter in size. The tip is inserted into a hole drilled through the center of the PI part ((column, row) = (20, 20)) to illuminate the sample. To map the pressure field, a hydrophone was mounted on two orthogonal linear translation stages. By stepping the hydrophone to the positions above those through holes in the NI part and recording the pressure amplitude at every step, we acquired the spatial field distribution of the ultrasound wave focusing pattern.

Fig.2 (a) shows the pressure field map in the NI part at 60.5 KHz with the PI/NI interface along x=0. The pressure amplitude is normalized to unity. A tight spot is observed in experiment as is evident from the plot. The pressure cross the focal plane along y direction is plotted in Fig. 2(c). The full width at half maximum (FWHM) was found to be 12.2mm, corresponding to a resolution of 0.5 wavelength in water.

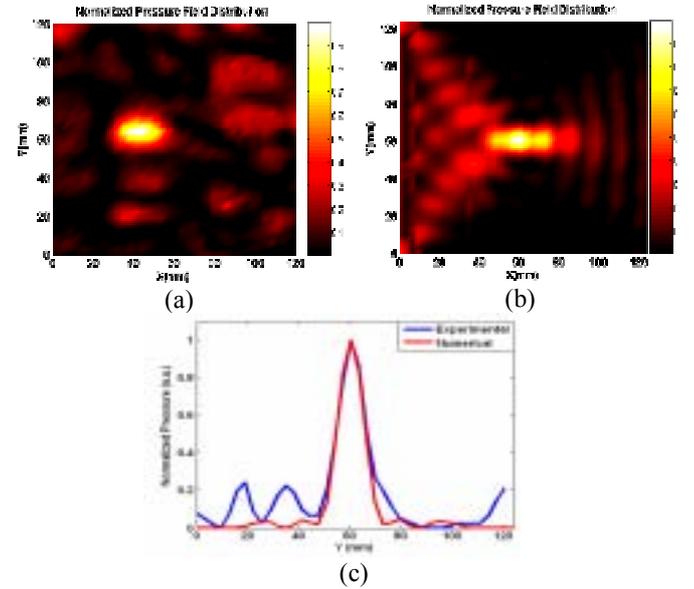

(a)  (b)

(c)

FIG. 2 (color). Pseudo colormap of the normalized pressure field distribution at 60.5 KHz. (a) Measured and (b) simulated field map of the acoustic NI metamaterial and (c) Line plot of pressure field cross the focal plane parallel to interface

For numerical verification, lumped circuit simulation of this acoustic network was performed by using commercial circuit simulator SPICE. Comparison of Fig. 2 (a) and (b) shows that the field plots found through simulation is in remarkable agreement with the experimental results. In Fig. 2(c), the measured data in blue line is shifted to left by 3.175 mm for comparison purpose. The comparison demonstrates a very good match in the focus width between the measurement and the numerical simulation. We also plotted the full width at half maximum (FWHM) at different frequencies in Fig. 3(a).

The optimal focus imaging is observed at 60.5 KHz from both experimental and numerical results.

The focal length defined as the distance between focus and PI/NI interface is plotted in Fig.3 (b) as a function of frequency. Ray acoustics is utilized to estimate the focal length as shown in the blue solid line. The magnitude of the refractive index in the acoustic metamaterial decreases from 1.19 to 0.85 as the frequency increases from 56 to 66 KHz. And this analysis predicts that the negative refractive index approaches -1 relative to the PI part at 60.5 KHz. The decrease of the index magnitude over this frequency range causes the focal length decreasing from 79.27 to 37.6mm. The lumped circuit simulation gives the dashed green curve while the red stars show the measurement data. The three curves present a good match in trend. However, around 10 mm shift is observed and we are investigating this discrepancy.

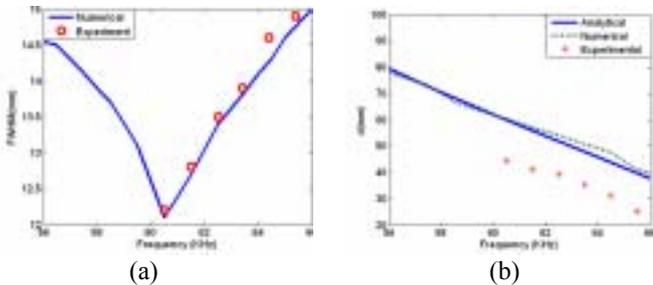

(a)          (b)

FIG. 3 (color). (a) Measured and calculated FWHM of the focus as a function of frequency. Blue solid line is calculated using acoustic circuit model and red circles represent experimental data. (b) Measured and calculated focal length as a function of frequency. The analytical curve (blue solid line) is based on ray propagation and the green dashed line is numerical results from acoustic circuit model. Experimentally obtained data are shown by red circles.

In order to achieve high-quality focus imaging, the ratio of refractive index should be -1 at the PI/NI interface. Only when the index is matched, based on ray acoustics, the angle of refraction equals the angle of incidence for each ray such that all rays can be brought to the same focal spot in the NI part. However, due to the loss and variation of the inductance and capacitance from their design values resulted from machining tolerance; the refractive index is not exactly matched in the measurement. Therefore, the distance between the focus and the interface varies for different incident angle as result of aberration [8]. Due to this index mismatch, we observed that the focal spot elongated along x direction while remain narrow along the direction parallel to the interface in the experiment. And the focus is in a position closer to the interface than the source. The best focusing resolution is observed at 60.5 KHz. We expect that the ratio of refractive index might approach -1 at lower frequency. However verification of this is beyond the operation frequency range of our transducer in the experiment. The slight material loss in the measurement also significantly degrades the focusing resolution as shown in several papers [26-27]. It was noted that single PI/NI interface does not allow the enough growth of evanescent fields to achieve sub diffraction focusing [8] while sandwich structure (two PI/NI interfaces) offers better chance to overcome the diffraction limit [10].

In summary, the emission of a point source at kilohertz frequency was brought to a focus through the PI/NI interface because of the negative refraction in this ultrasonic metamaterial, which is expected to be a step toward a novel acoustic imaging lens. The resolution of 0.5wavelength was recorded by focusing the acoustic field of a point source. This is not sub diffraction imaging, but among the best achievable passive acoustic imaging elements. The unit cell of the acoustic network is only one eighth of the operating wavelength, making the lens in a compact size. Compared with conventional lenses, the flat thin slab lens takes advantages in that there is no need to manufacture the shapes of spherical curvatures and the focus position is insensitive to the offset of source along the axis. Also this negative index lens offers tunable focal length at different frequencies. More generally, this design approach may lead to novel strategies of acoustic cloak for camouflage under sonar.

**Acknowledgements**
This work is partially supported by DARPA grant HR0011-05-3-0002. We are grateful to the inspiring private communication with Prof. Jianyi Xu [28] from Nanjing University, China at the initial stage of the sample design and the helpful discussion with Prof. William O'Brien at UIUC regarding experimental setups.